\documentclass[twocolumn,preprintnumbers,amsmath,amssymb]{revtex4}

\usepackage{graphicx}
\usepackage{dcolumn}
\usepackage{bm}

\input epsf

\begin{document}


\title{Quenching of the Haldane gap in LiVSi$_2$O$_6$ and related compounds}

\author{B. Pedrini\footnote[1]{Current address: \textit{The Scripps Research Institute, 10550 North Torrey Pines Road, La Jolla, CA 92037, USA}}, J. L. Gavilano\footnote[2]{Current address: \textit{Laboratory for Neutron Scattering, PSI, 5232 Villigen PSI, Switzerland}}, H. R. Ott, S. M. Kazakov\footnote[3]{Current address: \textit{Crystal Chemistry Laboratory, Inorganic Chemistry Division, Chemical Departement, Moscow State University, Leninskie Gory, 119992 Moscow, Russia}} and J. Karpinski}
\affiliation{
Laboratorium f\"{u}r Festk\"{o}rperphysik, ETH-H\"{o}nggerberg,\\
CH-8093~Z\"{u}rich, Switzerland
}

\author{S. Wessel}
\affiliation{
Institut f\"{u}r Theoretische Physik III, Universit\"{a}t Stuttgart\\
Pfaffenvaldring 57, D-70550~Stuttgart, Germany\\
}

\date{\today}

\begin{abstract}
We report results of susceptibility $\chi$ and $^7$Li NMR measurements on LiVSi$_2$O$_6$.
The temperature dependence of the magnetic susceptibility $\chi(T)$ exhibits a broad maximum, typical for low-dimensional magnetic systems.
Quantitatively it is in agreement with the expectation for an $S=1$ spin chain, represented by the structural arrangement of V ions.
The NMR results indicate antiferromagnetic ordering below $T_{\mathrm{N}}=24$ K.
The intra- and interchain coupling $J$ and $J_{\mathrm{p}}$ for LiVSi$_2$O$_6$, and also for its sister compounds LiVGe$_2$O$_6$, NaVSi$_2$O$_6$ and 
NaVGe$_2$O$_6$, are obtained via a modified random phase approximation which takes into account results of quantum Monte Carlo calculations.
While $J_{\mathrm{p}}$ is almost constant across the series, $J$ varies by a factor of 5, decreasing with increasing lattice constant along the chain direction.
The comparison between experimental and theoretical susceptibility data suggests the presence of an easy-axis magnetic anisotropy, which explains the formation of an energy gap in the magnetic excitation spectrum below $T_{\mathrm{N}}$, indicated by the variation of the NMR spin-lattice relaxation rate at $T\ll T_{\mathrm{N}}$.
\end{abstract}


\maketitle

\section{\label{SecIntroduction}Introduction}

In compounds of the series AVX$_2$O$_6$, where A=Li,Na and X=Si,Ge, the trivalent V ions occupy the centers of VO$_6$ octahedra forming chain-like structured elements.
Because of this arrangement of the $S=1$ magnetic moments , these materials may in principle be considered as a physical realization of an ensemble of one-dimensional spin $S=1$ chains with an antiferromagnetic intrachain coupling $J$.
These, according to Haldane ~\cite{Haldane_1983_PhysLett}, are expected to adopt a non magnetic ground state, separated by an energy gap from magnetically excited states.

Nevertheless the measured magnetic susceptibility $\chi$ of the first member of this family, LiVGe$_2$O$_6$ \cite{Millet_1999_PRL}, was found to be clearly different from the expected $\chi(T)$ of spin $S=1$ chains.
The discrepancy was first attributed to some sophisticated types of intrachain interactions \cite{Millet_1999_PRL,Gavilano_2000_PRL}, but subsequent $^7$Li NMR experiments \cite{Vonlanthen_2002_PRB} revealed a three-dimensional antiferromagnetic order below $T_{\mathrm{N}}=24$ K, most likely due to a small but non negligible coupling $J_{\mathrm{p}}$  between moments of neighbouring chains.
Magnetic susceptibility and $^{23}$Na NMR measurements later confirmed a similar behaviour of NaVGe$_2$O$_6$ with $T_{\mathrm{N}}=18$ K \cite{Pedrini_2004_PRB}.

It is conceivable that the compounds of the series AVX$_2$O$_6$ represent a physical realization of solids with spin $S=1$ chains but with interchain interactions that are large enough to quench the expected Haldane gap and to provoke a magnetically ordered ground state, as was suggested in Ref. \cite{Vasiliev_2004_PRB} on the basis of magnetic susceptibility and specific heat measurements.

In this article, we present the results of measurements of the magnetic susceptibility and of the $^{9}$Li NMR response of LiVSi$_2$O$_6$.
Inspecting the temperature dependence of the magnetic susceptibility reveals no anomaly that would indicate the onset of magnetic order.
However,  an abrupt broadening of the NMR spectra upon cooling reveals a transition from the high-temperature paramagnetic to the low-temperature antiferromagnetically ordered phase at $T_{\mathrm{N}}=24$ K.
In addition, a prominent peak in the temperature dependence of the NMR spin-lattice relaxation rate $T_1^{-1}(T)$ at the same temperature supports this interpretation.

Including previously obtained experimental data we extend our discussion to cover the series AVX$_2$O$_6$. 
Using a modified random phase approximation (RPA), we determine the values of $J$ and $J_{\mathrm{p}}$ from the corresponding values of $T_{\mathrm{N}}$ and 
$T_{\mathrm{max}}$, where the latter temperature corresponds to the maximum of the magnetic susceptibility.
While $J_{\mathrm{p}}$ is approximately constant, $J$ varies by a factor of 5 across the AVX$_2$O$_6$ series.
We then compare the measured magnetic susceptibilities to those calculated by quantum Monte Carlo (QMC) methods with corresponding 
values of $J$ and $J_{\mathrm{p}}$, and 
argue that the observed discrepancies may be explained by a single ion anisotropy.

The paper is organized as follows.
In Section \ref{SecSampleCrystStruct} we address the crystal structure and the synthesis of the LiVSi$_2$O$_6$ sample.
Section \ref{SecSusceptibility} is devoted to presenting the results of magnetic susceptibility measurements, and in Section \ref{SecNMR} we display the results of the NMR measurents.
In Section \ref{SecJpJ} we determine $J$ and $J_{\mathrm{p}}$ for the compounds in the AVX$_2$O$_6$ series, and compare the experimental and calculated magnetic susceptibilities.
In the Appendix we present the technical aspects of the QMC calculations used in this paper in some detail.

\section{\label{SecSampleCrystStruct}Crystal Structure and Sample}

\begin{figure}[t]
 \begin{center}
  \leavevmode
  \epsfxsize=1.0\columnwidth \epsfbox{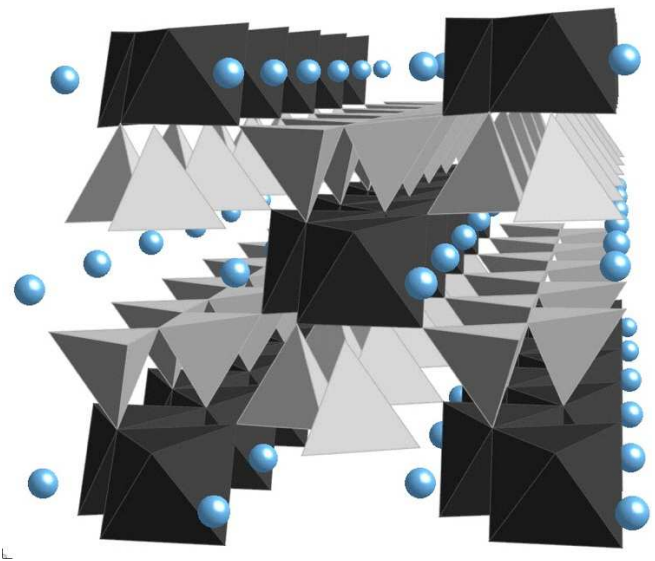}
\caption{
Representation of the crystal structure of LiVSi$_2$O$_6$.
The V$^{3+}$ ions are located in the center of the dark-grey VO$_6$ octahedra, while the Si$^{4+}$ ions occupy the centers of the light-grey SiO$_4$ tetrahedra.
The Li$^+$ions are represented by the spheres. 
}
\label{FigStructure}
\end{center}
\end{figure}

The crystal structure of LiVSi$_2$O$_6$ is schematically shown in Fig. \ref{FigStructure}.
Typical for the AVX$_2$O$_6$ series, we note the chains of isolated VO$_6$ octahedra joined at the edges.
These chains are linked and kept apart by SiO$_4$ tetrahedra.
Considering the most likely oxidation states of Li$^+$, Si$^{4+}$ and O$^{2-}$, the V ions are expected to be trivalent.
The magnetic moments are thus due to two electrons occupying the 3$d$ shell of V$^{3+}$ ions, thus forming an $S=1$ configuration.
The effective interaction between these moments is mediated by the oxygen $2p$-electrons, and is direct only for neighbours placed on the same chain.
In this geometry, the intrachain nearest neigbour coupling $J$ is expected to be much larger than any possible coupling $J_{\mathrm{p}}$ between moments on different chains.

LiVSi$_2$O$_6$ crystallizes in a monoclinic structure, space group $C2/c$ \cite{Satto_1997_ActaCrystC}.
The LiVSi$_2$O$_6$ powder sample was prepared with the following procedure.
First, a precursor with nominal composition LiSi$_2$Ox was synthesized by heating a mixture of Li$_2$CO$_3$ (Aldrich, 99.99\%) and SiO$_2$ (Aldrich, 99.995\%) at 725$^{\circ}$C for 50 hours in air and a reannealing at 730$^{\circ}$C for 20 hours in air. 
Then a stoichiometric mixture of LiSi$_2$Ox and V$_2$O$_3$ (Aldrich,99.99\%) was pressed into pellets and annealed at 990$^{\circ}$C for 40 hours in an evacuated silica tube. 
Subsequent heat treatment for 60 hours in another evacuated silica tube was necessary to obtain the desired composition of LiVSi$_2$O$_6$.
According to x-ray powder diffraction data, the sample contained 95\% of the desired phase with the pyroxene structure and 5 mass percent of non-magnetic SiO$_2$.
The lattice parameters of LiVSi$_2$O$_6$ were determined by room temperature X-ray diffraction measurements, and resulted in $a=9.634$ \AA, $b=8.586$ \AA, $c=5.304$ \AA, and $\beta=109.69^{\circ}$.

In Table \ref{TabCrystStruct} we list the structural parameters of the unit cell of the compounds in the AVX$_2$O$_6$ series.
In Table \ref{TabVVBondData} we also list the length of the intrachain V-V bonds, as well as the angle between two successive V-V bonds in the same chain.
These values were obtained by using a V-V bond corresponding to the vector 
$\vec{v}=\delta a\vec{a}+\delta b\vec{b}+\delta c\vec{c}$, where $\vec{a}$, $\vec{b}$ and $\vec{c}$ are the unit cell vectors.
From the geometry of the chains it follows that $\delta a=0$, since the chains lie in $a$-planes, $\delta c=0.5$ and $\delta b=\pm|\delta b|$, where the sign switches because of the zig-zag character of the chains.
The parameter $|\delta b|$ is different for the individual compounds, and is calculated from the exact positions of the V atoms.
We note only small variations of the V-V bond length and the V-V-V angle across the series.

\begin{table}[t]
 \begin{center}
 \begin{tabular}{|l|c|c|c|c|c|}
   \hline
    & & & & & \\
    & Space group & $a$ (\AA) & $b$ (\AA) & $c$ (\AA) & $\beta$ \\
    & & & & & \\
    \hline
    & & & & & \\
    LiVSi$_2$O$_6$
      & $C2/c$ & 9.634 & 8.586 & 5.304 & 109.69$^{\circ}$ \\
    NaVSi$_2$O$_6$ \cite{Vasiliev_2004_PRB}
      & $C2/c$ & 9.634 & 8.741 & 5.296 & 109.90$^{\circ}$ \\
    LiVGe$_2$O$_6$ \cite{Millet_1999_PRL}
      & $P2_1/c$ & 9.863 & 8.763 & 5.409 & 108.21$^{\circ}$ \\
    NaVGe$_2$O$_6$ \cite{Pedrini_2004_PRB}
      & $P2_1/c$ & 9.960 & 8.844 & 5.486 & 106.50$^{\circ}$ \\
    & & & & & \\
    \hline
 \end{tabular}
\caption{
Values of the unit cell parameters in the AVX$_2$O$_6$ family.}
\label{TabCrystStruct}
\end{center}
\end{table}

\begin{table}[t]
 \begin{center}
 \begin{tabular}{|l|c|c|c|}
   \hline
    & & &  \\
    & $\delta b$ (\AA) & V-V length (\AA) & V-V-V angle \\
    & & & \\
    \hline
    & & & \\
    LiVSi$_2$O$_6$
    & 0.188 & 3.105 & 62.7$^{\circ}$\\
    LiVGe$_2$O$_6$
    & 0.186 & 3.158 & 62.1$^{\circ}$\\
    NaVGe$_2$O$_6$
    & 0.193 & 3.256 & 63.2$^{\circ}$\\
    & & & \\
    \hline
 \end{tabular}
\caption{
V-V bond data for selected compounds in the AVX$_2$O$_6$ family.}
\label{TabVVBondData}
\end{center}
\end{table}

\section{\label{SecSusceptibility}Magnetic Susceptibility}

Using a commercial SQUID magnetometer, we measured the magnetization $M$ of 18.8 mg of powdered LiVSi$_2$O$_6$ (number of mols $N=8.95\times 10^{-5}$), for temperatures $T$ between 4 and 340 K and in magnetic fields $\mu_0H$ between 0.01 T and 5 T.
Apart from a temperature independent component, the molar magnetic susceptibility $\chi=M/(H N)$ was found to be identical in different magnetic fields above 10 K.

\begin{figure}[t]
 \begin{center}
  \leavevmode
  \epsfxsize=1.0\columnwidth \epsfbox{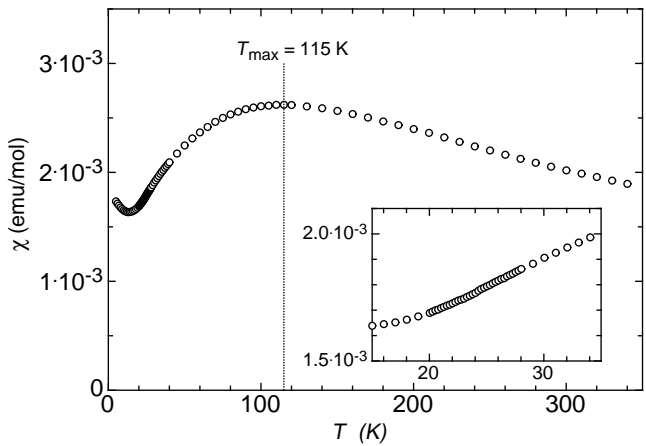}
\caption{
Magnetic susceptibility $\chi$ of LiVSi$_2$O$_6$ as a function of temperature $T$, measured in a magnetic field $H=5$ T.
The inset magnifies the temperature range 15-35 K.}
\label{FigChiTLVSO}
\end{center}
\end{figure}

In Fig. \ref{FigChiTLVSO} we display an example of $\chi(T)$, measured in a magnetic field $\mu_0H=5$ T.
The temperatures are too low to identify a high-temperature Curie-Weiss behaviour and thus to establish the oxidation state of the V ions.
Reducing the temperature from 340 K, $\chi(T)$ increases and reaches a broad maximum at $T_{\mathrm{max}}=115\pm1$ K.
This type of temperature dependence of $\chi$ is typical for low-dimensional magnetic systems.
Upon cooling to below 15 K, $\chi(T)$ exhibits a smooth upturn, which is attributed to the low-temperature Curie-Weiss tail caused by a small amount of paramagnetic impurities.

We note that in this data set there is no evidence for any anomalies in $\chi(T)$ at temperatures below $T_{\mathrm{max}}$.
In particular, as is shown in the inset of Fig. \ref{FigChiTLVSO}, $\chi(T)$ varies smoothly between 20 K and 30 K, the temperature range in which, as reported in Sec. \ref{SecNMR}, a magnetic transition is revealed by the NMR data.

This is somewhat surprising because for LiVGe$_2$O$_6$ \cite{Gavilano_2000_PRL}, NaVGe$_2$O$_6$ \cite{Pedrini_2004_PRB} and NaVSi$_2$O$_6$ \cite{Vasiliev_2004_PRB}, a prominent kink in $\chi(T)$ indicates the onset of a three-dimensionally antiferromagnetically ordered state upon cooling.

\section{\label{SecNMR}$^7\mathrm{Li}$ NMR}

The NMR experiments probed the same sample as that used for the magnetization measurements.
The $^7$Li NMR spectra were obtained by monitoring the integrated spin-echo intensity as a function of the irradiation frequency $f$ in a fixed external magnetic field $H$.
The selected values $\mu_0H=7.0495$ T ($7.05$ T in the following) and $\mu_0H=4.5533$ T ($4.55$ T in the following) were calibrated by monitoring the $^2$D resonance frequency in liquid D$_2$O.
The respective $^7$Li reference frequencies, $f_0=116.643$ MHz and $f_0=75.340$ MHz, were obtained from $f_0=\gamma\mu_0 H$, with $\gamma=16.5463$ s$^{-1}$T$^{-1}$ as the gyromagnetic factor of the $^7$Li nuclei. 
The spin echo was generated  with a two pulse $\pi/2$-delay-$\pi$ spin-echo sequence.

\begin{figure}[t]
 \begin{center}
  \leavevmode
  \epsfxsize=1.0\columnwidth \epsfbox{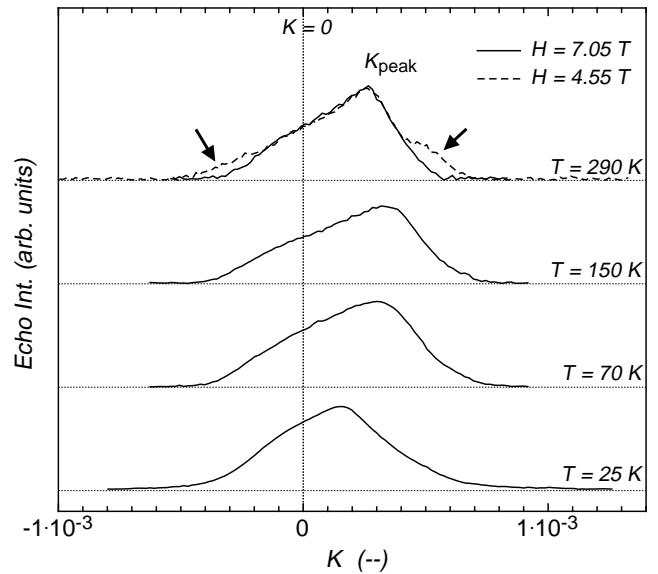}
\caption{
 $^7$Li NMR spectra of LiVSi$_2$O$_6$ for selected temperatures above 25 K. 
The integrated echo intensity, multiplied by the temperature $T$, is plotted as a function of the frequency shift $K$.
In the upper graph, two spectra taken at $T=290$ K but in different magnetic fields $\mu_0H=4.55$ T and $\mu_0H=7.05$ T are compared. 
}
\label{FigSpecParam}
\end{center}
\end{figure}

Fig. \ref{FigSpecParam} shows examples of recorded spectra for $\mu_0H=7.05$ T and $\mu_0H=4.55$ T at selected temperatures above 25 K.
They were obtained by irradiating a frequency window of about 20 kHz.
The integrated echo intensity, multiplied by the temperature $T$, is displayed as a function of the frequency shift
\begin{equation}
  K=\frac{f}{f_0}-1
  \quad,
\end{equation}
where $f$ is the irradiation frequency.
In the temperature range between 25 K and 294 K the shape and the total intensity of the NMR signal are essentially the same .

For $T=293$ K, we show two spectra measured in the two different magnetic fields of 7.05 T and 4.55 T.
As shown in Fig.\ref{FigSpecParam}, they coincide in their central parts but exhibit some differences, indicated by the arrows, at the upper and lower ends of the signal.
This is attributed to the first order quadrupolar splitting of the nuclear transitions between the Zeeman states $m=\pm3/2$ and $ m=\pm1/2$ \cite{Carter}. This frequency splitting is field independent and, in the spectra in the higher field, is masked by the magnetic broadening of the NMR line, which is proportional to the applied magnetic field.
The width in frequency space at the basis of the spectrum for $\mu_0H=4.55$ T allows for an estimate of the upper limit of the quadrupolar frequency  $f_Q\leq80$ kHz of the $^7$Li nuclei.

\begin{figure}[t]
 \begin{center}
  \leavevmode
  \epsfxsize=1.0\columnwidth \epsfbox{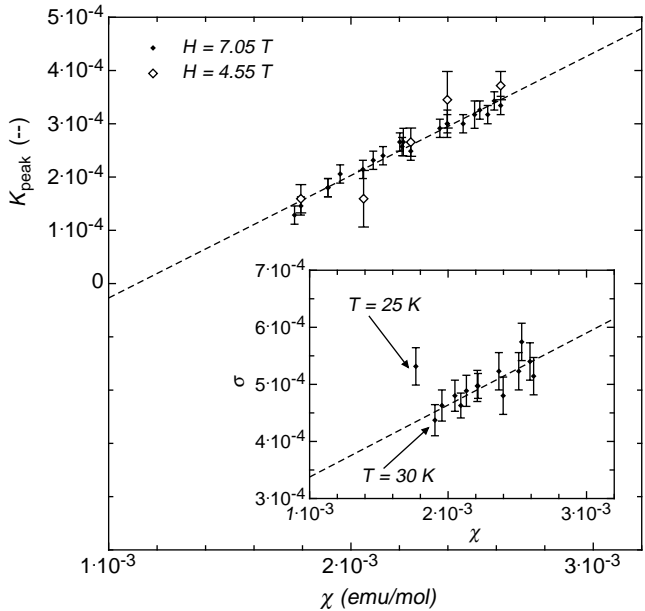}
\caption{
Relative frequency shift $K_{\mathrm{peak}}$ of the peak frequency $f_{\mathrm{peak}}$ of the $^7$Li NMR line, displayed as a function of the magnetic susceptibility $\chi$. The data are for temperatures $T>25$ K and in magnetic fields $H=7.05$ T and $H=4.55$ T.
The dashed line represents the best linear fit to the experimental data.
Inset:
The dashed line again represents the best linear fit.
}
\label{FigKChi}
\end{center}
\end{figure}

Above 25 K the lineshape of the signal is compatible with that of a randomly oriented powder with anisotropic shifts.
The anisotropic part may be attributed to the dipolar coupling between the Li nuclei and the paramagnetic V-moments.
Such lineshapes were already observed in the case of NaVGe$_2$O$_6$ \cite{Pedrini_2004_PRB} and LiVGe$_2$O$_6$ \cite{Vonlanthen_2002_PRB}.

In Fig. \ref{FigKChi}, the $y$-axis corresponds to the shift $K_{\mathrm{peak}}$ of the frequency $f_{\mathrm{peak}}$ at the maximum of the NMR signal.
For $T\geq25$ K, $K_{\mathrm{peak}}$ is plotted as a function of $\chi(T)$.
The data for the two different fields collapse onto the same curve, and $K_{\mathrm{peak}}(\chi)$ can reasonably well be fitted by
\begin{equation}
  K_{\mathrm{peak}}=a_{\mathrm{peak}}\chi+c_{\mathrm{peak}}
  \quad,
\end{equation}
indicating that the temperature dependence of the shift of the NMR signal is of purely magnetic origin.
The hyperfine coupling corresponding to the signal peak is \cite{Carter}
\begin{equation}
  A_{\mathrm{peak}}=a_{\mathrm{peak}}\cdot N\mu_B = 1280\pm50\;\mathrm{G}
  \quad.
\end{equation}
If the hyperfine coupling expected for a purely dipolar interaction of the V moments with the Li nuclei is calculated as in Refs. \cite{Vonlanthen_2002_PRB,Pedrini_2004_PRB}, the result is $A_{\mathrm{peak}}^\mathrm{{dip}}=210$ G.
The discrepancy between this calculated and the observed value is attributed to an additional non-dipolar hyperfine coupling $A_0=1070\pm50$ G, such that $A_{\mathrm{peak}}=A_{\mathrm{peak}}^\mathrm{{dip}}+A_0$.
In Table \ref{TabHypCoup} we compare the values of $A_{\mathrm{peak}}$ and $A_0$ for three compounds of the AVX$_2$O$_6$ series.

In the inset of Fig. \ref{FigKChi} we plot 
\begin{equation}
  \sigma=\frac{\mathrm{FWHM}}{f_0}
  \quad,
\end{equation}
where FWHM is the full width at half maximum of the NMR signal and $f_0$ the above mentioned reference frequency, again as a function of $\chi$.
Above 30 K, the data can be accomodated by 
\begin{equation}
  \label{EqSigmaChi}
  \sigma=a_{\sigma}\chi+c_{\sigma}
  \quad.
\end{equation}
The corresponding anisotropic part of the hyperfine coupling
\begin{equation}
  \sigma_A=a_{\sigma}\cdot N\mu_B = 700\pm100\;\mathrm{G}
  \quad,
\end{equation}
is close to the value $\sigma_A^\mathrm{{dip}}=650$ G if calculated in the same way as in Ref. \cite{Pedrini_2004_PRB}, assuming purely dipolar coupling between the V magnetic moments and the Li nuclei.

\begin{table}[t]
 \begin{center}
 \begin{tabular}{|l|c|c|}
   \hline
    & &  \\
    & $A_{\mathrm{peak}}$ (G)& $A_0$ (G) \\
    & & \\
    \hline
    & &  \\
    LiVSi$_2$O$_6$
      & $1280\pm50$ & $1050\pm50$ \\
    LiVGe$_2$O$_6$ \cite{Vonlanthen_2002_PRB}
      & $580\;\;\;\;\;\;$ & $480\;\;\;\;\;\;$ \\
    NaVGe$_2$O$_6$ \cite{Pedrini_2004_PRB}
      & $1300\pm50$ & $1140\pm50$ \\
    & & \\
    \hline
 \end{tabular}
\caption{
Values of $A_{\mathrm{peak}}$ and $A_0$ for selected compounds in the AVX$_2$O$_6$ family.
}
\label{TabHypCoup}
\end{center}
\end{table}

\begin{figure}[t]
 \begin{center}
  \leavevmode
  \epsfxsize=1.0\columnwidth \epsfbox{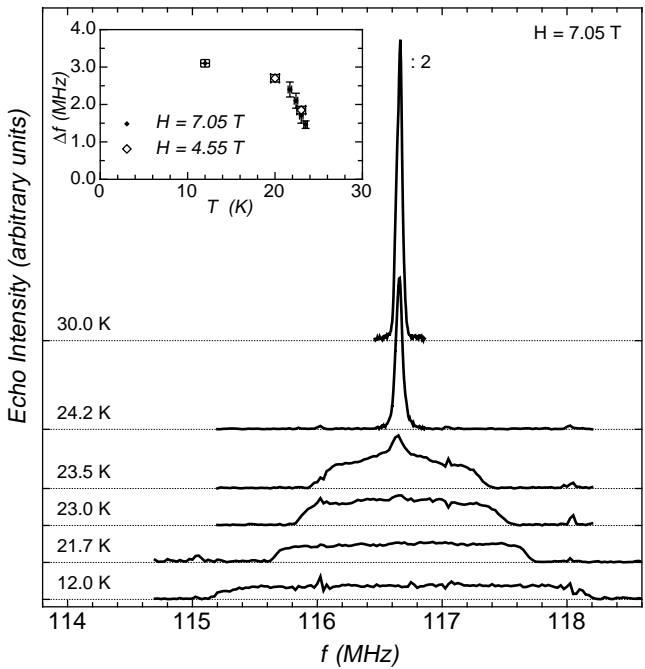}
\caption{
$^7$Li NMR spectra of LiVSi$_2$O$_6$ for selected temperatures below 25 K. 
The echo intensity is plotted as a function of the frequency $f$ for selected temperatures below 25 K.
The measurement were performed in a fixed magnetic field $H=7.05$ T.
Inset: line width $\Delta f$ as a function of $T$, in magnetic fields $H=4.55$ T and $H=7.05$ T.
}
\label{FigSpecPhTr}
\end{center}
\end{figure}

The inset of Fig. \ref{FigKChi} also shows that the value of $\sigma$ at 25 K is distinctly larger than predicted by Eq. (\ref{EqSigmaChi}), thus indicating the onset of antiferromagnetic correlations.
For a discussion of the low-temperature behaviour we turn our attention to the NMR spectra at temperatures below 25 K.
Fig. \ref{FigSpecPhTr} displays examples of spectra measured at temperatures below 30 K and in a magnetic field $\mu_0H=7.05$ T, recorded by irradiating a frequency window of about 100 kHz.
Above $T_{\mathrm{N}}=24$ K, a single NMR line characterizing the paramagnetic state is observed.
Upon cooling its intensity decreases drastically between 25 K and 24 K, and the narrow line has completely vanished at 23 K.
Below $T_{\mathrm{N}}=24$ K, a broad, rectangular shaped NMR signal develops with decreasing temperature.
This shape is typical for ordered moments in a powder sample.
The interpretation of an onset of magnetic order is also substantiated by the $H$- and $T$-dependence of the width $\Delta f$ of the signal, which is represented as a function of temperature $T$ in the inset of Fig. \ref{FigSpecPhTr} for magnetic fields $\mu_0H=4.55$ T and $\mu_0H=7.05$ T.
Indeed, $\Delta f$ is field independent and, for $T\rightarrow0$, we argue that $\Delta f$ tends to saturate at a constant value of $\Delta f_{\ast}=3.4\pm0.4$ MHz.
Thus, below $T_{\mathrm{N}}$, the magnetic moments residing on the $V$ ions adopt a three-dimensional antiferromagnetic order. The internal magnetic field $H_{\mathrm{int}}$ at the Li sites attains a saturation value $H_{\mathrm{int}}=\Delta f_{\ast}/2\gamma=1030\pm60$ G.

\begin{figure}[t]
 \begin{center}
  \leavevmode
  \epsfxsize=1.0\columnwidth \epsfbox{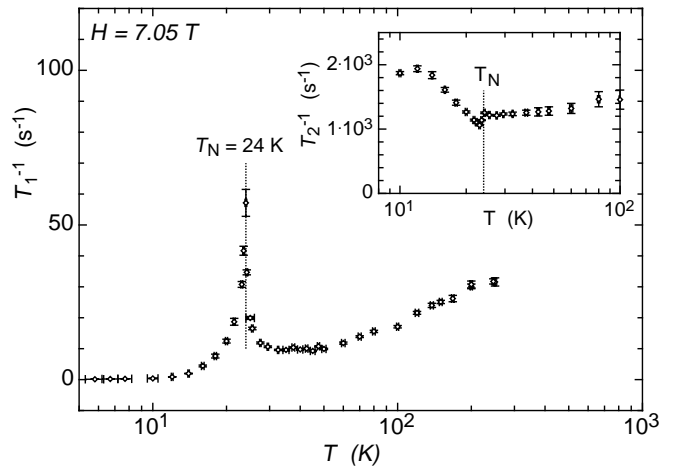}
\caption{
$^7$Li spin-lattice relaxation rate $T_1^{-1}$ of LiVSi$_2$O$_6$ as a function of temperature $T$.
The maximum of $T_1^{-1}$ at $T_{\mathrm{N}}$ is indicated by the dotted vertical line.
Inset: $^7$Li spin-spin relaxation rate $T_2^{-1}$ as a function of $T$. 
$T_{\mathrm{N}}$ is indicated by the dotted vertical line.
}
\label{FigT1T2}
\end{center}
\end{figure}

To gain further information on the phase transition at $T_{\mathrm{N}}=24$ K, we measured the temperature dependence of the spin-lattice relaxation rate, $T_1^{-1}(T)$.
$T_1^{-1}$ was determined by monitoring the recovery of the $^7$Li nuclear magnetization $m$ after the application of a long comb of radiofrequency pulses.
The experiments were performed in a magnetic field $\mu_0H=7.05$ T, irradiating a frequency window of about 100 kHz in the center of the NMR signal.
An exponential recovery
\begin{equation}
  m(t)=m_{\infty}[1-\exp(-t/T_1)]
\end{equation}
with recovery time $t$  was observed across the entire temperature range between 5 and 295 K.
The single exponential law is appropriate, since we simultaneously irradiate all the three possible transitions of the $I=3/2$ Li nuclear spins, with the maximum splitting of the order of $f_{\mathrm{Q}}\approx80$ kHz.
In Fig. \ref{FigT1T2} we display the temperature dependence of the spin-lattice relaxation rate $T_1^{-1}$.
Above $T_{\mathrm{N}}$, $T_1^{-1}$ slowly decreases by a factor of three between 295 K and 30 K upon cooling.
A prominent peak in $T_1^{-1}(T)$ at $T_{\mathrm{N}}=24$ K reflects the magnetic phase transition.

\begin{figure}[t]
 \begin{center}
  \leavevmode
  \epsfxsize=1.0\columnwidth \epsfbox{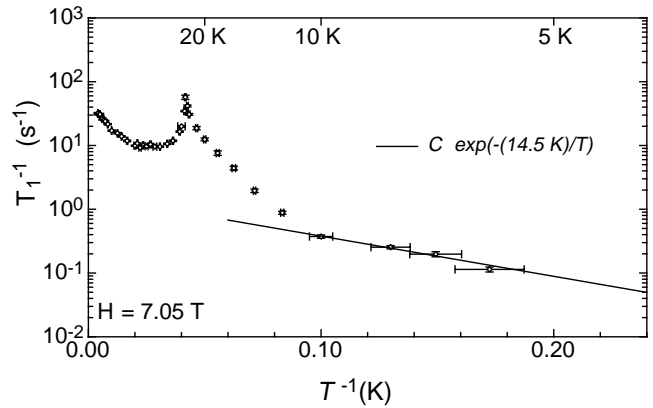}
\caption{
$^7$Li spin-lattice relaxation rate $T_1^{-1}$ of LiVSi$_2$O$_6$ as a function of inverse temperature $T^{-1}$.
The solid line represents the function $T_1^{-1}=C\exp(-\Delta/T)$ with $\Delta=14.5$ K.
}
\label{FigT1LowT}
\end{center}
\end{figure}

The temperature dependence of the spin-lattice relaxation rate at low temperature provides additional information about the low-energy magnetic excitations.
Below $T_{\mathrm{N}}$, $T_1^{-1}$ decreases by orders of magnitude (a factor of 100 between 25 K and 10 K, as shown in Fig. \ref{FigT1LowT}).
indicating the formation of a gap in the spectrum of magnon excitations in the antiferromagnetically ordered phase.
This interpretation is supported by $T_1^{-1}(T)$ well below $T_{\mathrm{N}}$, which varies according to \cite{Barak_1974_PRB}
\begin{equation}
  T_1^{-1}(T)\sim\exp(-\Delta/T)
  \quad,
\end{equation}
as is emphasized by the solid line in Fig. \ref{FigT1LowT}.
Because of the presence of a small amount of paramagnetic impurities, the actual value of the gap may be somewhat larger than the  estimated value $\Delta=14.5$ K.
A comparison of gap values, all established from $T_1^{-1}(T)$ well below $T_{\mathrm{N}}$, is made in Table \ref{TabT1Gap} for three AVX$_2$O$_6$ compounds; they all turn out to be almost equal.
Such gaps are usually attributed to an easy-axis single-ion magnetic anisotropy \cite{Hone_1969_PR}, an aspect discussed in more detail in Sec. \ref{SecJpJ}.

\begin{table}[t]
 \begin{center}
 \begin{tabular}{|l|c|}
   \hline
    &   \\
    & $\Delta$ (K) \\
    &  \\
    \hline
    &   \\
    LiVSi$_2$O$_6$
      & 14.5 \\
    LiVGe$_2$O$_6$ \cite{Vonlanthen_2002_PRB}
      & 14.7   \\
    NaVGe$_2$O$_6$ \cite{Pedrini_2004_PRB}
      & 12.5 \\
    &  \\
    \hline
 \end{tabular}
\caption{
Values of the gap in the magnetic excitation spectrum $\Delta$, calculated from the low-temperature dependence of $T_1^{-1}$, for selected compounds in the AVX$_2$O$_6$ family.
}
\label{TabT1Gap}
\end{center}
\end{table}

Finally, we consider the temperature dependence of the spin-spin relaxation rate $T_2^{-1}$.
The corresponding experiments were made under the same conditions as the $T_1$ experiments and $T_2$ was determined from fits to the decay of the echo intensity $m$ with the exponential law
\begin{equation}
  \label{EqT2Rec}
  m(\tilde{t})=m_{0}\exp(-\tilde{t}/T_2)
  \quad.
\end{equation}
Here $\tilde{t}=2\tau$ and $\tau$ is the delay between the $\pi/2$- and $\pi$-pulses which generate the spin echo.

In the paramagnetic phase above $T_{\mathrm{N}}$, $T_2^{-1}$ is approximately constant at a value $\approx(1.4\pm0.2)\times10^3$ s$^{-1}$.
As is shown in the inset of Fig. \ref{FigT1T2} the phase transition is revealed by a sudden reduction of $T_2^{-1}(T)$ at $T_{\mathrm{N}}$ upon cooling.
This decrease is followed by an upturn at even lower temperatures.
As $T\rightarrow0$, $T_2^{-1}$ tends to saturate towards the constant value $T_{2,T\rightarrow0}^{-1}=(1.9\pm0.1)\times10^3$.
This particular temperature dependence of $T_2^{-1}$ seems to be typical for the the AVX$_2$O$_6$ series (see the $T_2^{-1}(T)$ data in \cite{Pedrini_2004_PRB,Vonlanthen_2002_PRB}).
In Table \ref{TabT2Data} we display the values of $T_{2,T\rightarrow0}^{-1}$ and $T_{2,T\rightarrow\infty}^{-1}$, and notice that the values are of the same order of magnitude.

\begin{table}[t]
 \begin{center}
 \begin{tabular}{|l|c|c|}
   \hline
    & &  \\
    & $T_{2,T\rightarrow0}^{-1}$ ($10^3$ s$^{-1}$) & $T_{2,T\rightarrow\infty}^{-1}$ ($10^3$ s$^{-1}$) \\
    & & \\
    \hline
    & &  \\
    LiVSi$_2$O$_6$
      & $1.9\pm0.1$ & $1.4\pm0.2$ \\
    LiVGe$_2$O$_6$ \cite{Vonlanthen_2002_PRB}
      & $2.2\pm0.2$ & $1.2\pm0.2$ \\
    NaVGe$_2$O$_6$ \cite{Pedrini_2004_PRB}
      & $1.0\pm0.1$ & $0.6\pm0.1$ \\
    & & \\
    \hline
 \end{tabular}
\caption{
Values of $T_{2,T\rightarrow0}^{-1}$ and $T_{2,T\rightarrow\infty}^{-1}$ for selected compounds in the AVX$_2$O$_6$ family.
For the calculation of $T_2$ in LiVGe$_2$O$_6$, Eq. (\ref{EqT2Rec}) was replaced with $m(\tilde{t})=m_{0}\exp(-(\tilde{t}/T_2)^{1.4})$.
}
\label{TabT2Data}
\end{center}
\end{table}

In magnetically ordered states, $T_2^{-1}$ is often caused by magnon mediated interactions between nuclei. 
The unexpected upturn in $T_2^{-1}(T)$ upon cooling below $T_{\mathrm{N}}$ is incompatible with this scenario because the large gaps listed in Table \ref{TabT1Gap} would lead to a rapid decrease of $T_2^{-1}$.
No such decrease has been observed, in particular for LiVGe$_2$O$_6$ down to 1 K, as reported in Ref. \cite{Vonlanthen_2002_PRB}. Therefore it must be that other mechanisms are essential for the spin-spin relaxation.

\section{\label{SecJpJ}Weakly interacting spin $\mathrm{\bf{S=1}}$ chains in $\mathrm{\bf{(Li,Na)V(Ge,Si)}}_2\mathrm{\bf{O}}_6$.}

The results of the present and former studies \cite{Pedrini_2004_PRB,Vonlanthen_2002_PRB,Gavilano_2000_PRL} clearly indicate that in the (Li,Na)$_2$V(Ge,Si)$_2$O$_6$ series, the onset of the low temperature Haldane phase is intercepted by the developement of three-dimensional antiferromagnetic order.
It is known that even a very small interchain coupling may quench the Haldane gap and induce a three-dimensional antiferromagnetic order \cite{Koga_2000_PRB,Kim_2000_PRB,Yasuda_2005_PRL}.
In order to establish the ratios $J_{\mathrm{p}}/J$ between interchain and intrachain couling for the different compounds, we offer an analysis of the $\chi(T)$ 
data invoking quantum Monte Carlo calculations based on a model Hamiltonian.

We model the system of magnetic moments at the V sites by $S=1$ spins placed on a cubic lattice.
The coupling to the next neighbours along the chain direction is antiferromagnetic, $J>0$.
Each spin is allowed to interact with the four nearest spins of neighbouring chains (see Fig. \ref{FigStructure}), assuming an interchain coupling constant $J_{\mathrm{p}}$.
The spin  $S=1$ operators of the $i$-th spin on the $r$-chain are denoted by $\vec{S}_{r,i}$.
The Hamiltonian is the sum of two terms,
\begin{equation}
  \label{EqHamilton}
  H = H_J+H_{J_{\mathrm{p}}}
  \quad,
\end{equation}
where the first term represents the intrachain Hamiltonian for a collection of antiferromagnetic $S=1$ chains,
\begin{equation}
  \label{EqHamiltonJ}
   H_J = J\sum_{r;i}\vec{S}_{r,i}\cdot\vec{S}_{r,i+1}
   \quad (J>0) \quad.
\end{equation}
The second term in Eq. (\ref{EqHamilton}) is the interchain Hamiltonian
\begin{equation}
  \label{EqHamiltonJp}
   H_{J_{\mathrm{p}}} = J_{\mathrm{p}}\sum_{(r,s);i}\vec{S}_{r,i}\cdot\vec{S}_{s,i}
   \quad,
\end{equation}
where $(r,s)$ indicates a pair of neighbouring chains.

\begin{figure}[t]
 \begin{center}
  \leavevmode
  \epsfxsize=1.0\columnwidth \epsfbox{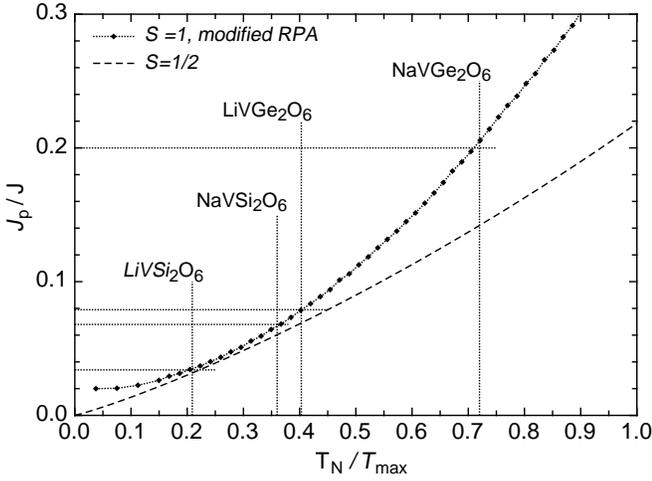}
\caption{
$J_{\mathrm{p}}/J$ as a function of $T_{\mathrm{N}}/T_{\mathrm{max}}$, resulting from the modified RPA approximation.
The value of $T_{\mathrm{N}}/T_{\mathrm{max}}$ for the AVX$_2$O$_6$ compounds are indicated.
The dotted line represents $J_{\mathrm{p}}/J(T_{\mathrm{N}}/T_{\mathrm{max}})$ for weakly interacting spin $S=1/2$ chains as given in Ref. \cite{Schulz_1996_PRL}.
Note the upper limit of $J_{\mathrm{p}}/J\sim0.02$ for $T_{\mathrm{max}}=0$.
}
\label{FigJpJTNTmax}
\end{center}
\end{figure}

In the random phase approximation (RPA)
\begin{equation}
  \label{EqJpJTNJ}
   J_{\mathrm{p}}/J=\frac{1}{4\xi\bar{\chi}^s(T_{\mathrm{N}}/J)}
   \quad,
\end{equation}
where $\bar{\chi}^s$ is the staggered susceptibility of an isolated $S=1$ chain ($J_{\mathrm{p}}=0$), which was calculated with QMC methods and reported in 
Ref. \cite{Pedrini_2004_PRB} (see also Appendix \ref{AppSusc}).
The factor 4 in the denominator corresponds to the number of neigbourhing chains.
The validity of Eq. (\ref{EqJpJTNJ}) was shown in Ref. \cite{Yasuda_2005_PRL} for $J_{\mathrm{p}}/J\leq0.2$ (see also Appendix \ref{AppSusc}), provided that the renormalized value of $\xi=0.695$, instead of the classically expected $\xi=1$, is used.
From $\chi(T)$ as obtained by QMC calculations, $T_{\mathrm{max}}/J$ may be expressed as a function of $J_{\mathrm{p}}/J$, using the approximation
\begin{equation}
  \label{EqJpJTmax}
  T_{\mathrm{max}}/J=a_2(J_{\mathrm{p}}/J)^2+a_1(J_{\mathrm{p}}/J) +a_0
  \quad,
\end{equation}
with $a_2=-0.95$, $a_1=0.69$ and $a_0=1.32$ (see Appendix \ref{AppSusc}).
Combining Eqs. (\ref{EqJpJTNJ}) and (\ref{EqJpJTmax}), the relation between
\begin{equation}
  \nonumber
  J_{\mathrm{p}}/J\;\mathrm{and}\; T_{\mathrm{N}}/T_{\mathrm{max}}
\end{equation}
may be established.
It is represented in Fig. \ref{FigJpJTNTmax} and allows to derive the values of $J_{\mathrm{p}}/J$ for the compounds of the series AVX$_2$O$_6$.
The results are listed in Table \ref{TabJpJ}.
These ratios differ from those given in Ref. \cite{Vasiliev_2004_PRB} because the latter were obtained using the formula given in Ref. \cite{Schulz_1996_PRL}, which is valid for spin $S=1/2$ chains, and is represented by the dotted line in Fig. \ref{FigJpJTNTmax}.

\begin{table}[t]
 \begin{center}
 \begin{tabular}{|l|c|c|c|c|c|}
   \hline
    & & & & & \\
    & $T_{\mathrm{max}}$ (K) & $T_{\mathrm{N}}$ (K) & $J_{\mathrm{p}}/J$ & $J/k_B$ (K) & $J_{\mathrm{p}}/k_B$ (K)\\
    & & & & & \\
    \hline
    & & & & & \\
    LiVSi$_2$O$_6$
      & $115\pm1$ & $24.0\pm0.5$ & $0.034$ & $85.7$ & $2.91$ \\
    NaVSi$_2$O$_6$
      & $48$ & $17.4$ & $0.068$ & $35.2$ & $2.40$ \\
    LiVGe$_2$O$_6$
      & $62\pm1$ & $25\pm0.5$ & $0.079$ & $45.3$ & $3.58$ \\
    NaVGe$_2$O$_6$ 
      & $25.0\pm0.5$ & $18.0\pm0.5$ & $0.20$ & $17.6$ & $3.54$ \\
    & & & & & \\
    \hline
 \end{tabular}
\caption{
List of magnetic parameters of the AVX$_2$O$_6$ family that are relevant in the comparison between calculations and experiments.}
\label{TabJpJ}
\end{center}
\end{table}

While the value of the interchain coupling $J_{\mathrm{p}}$ is almost constant across the series, the value of the intrachain coupling $J$ 
decreases by a factor of 5 from LiVSi$_2$O$_6$ to NaVSi$_2$O$_6$, LiVGe$_2$O$_6$ and NaVGe$_2$O$_6$.
The reason for such a variation cannot be established from the data at our disposal.
It does not appear to be explained by geometrical features of the V-chains alone, since the relevant parameters differ only by small amounts 
(0.5\% for the V-V bond length and $1^{\circ}$ for the V-V-V angle; see Table \ref{TabVVBondData}).
It may therefore be that the position of atoms other than V and the resulting electronic structure are relevant for the indirect mediation of the magnetic 
interaction between neighbouring V ions.
Inspecting the positions of the O ions around the V ions, it turns ou that $J$ decreases with increasing distortion of the VO$_6$-ochtahedra.

\begin{figure}[t]
 \begin{center}
  \leavevmode
  \epsfxsize=1.0\columnwidth \epsfbox{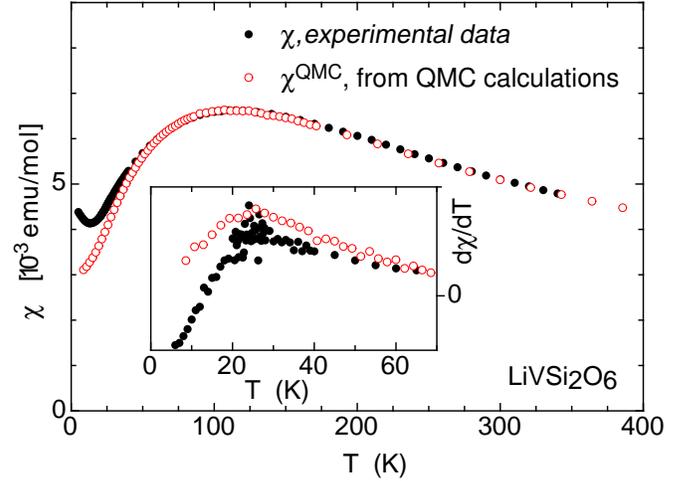}
\caption{
Magnetic susceptibility $\chi$ as a function of temperature $T$ of LiVSi$_2$O$_6$, compared with the $\chi(T)$ resulting from quantum Monte Carlo calculations.
Inset: $d\chi/dT(T)$.
Note that the noise in the experimental $d\chi/dT(T)$ data, computed from $\chi(T)$ represented in Fig. \ref{FigChiTLVSO}, does not allow to confirm or exclude the presence of a small anomaly like the one reported in Ref. \cite{Vasiliev_2004_PRB}.
}
\label{FigChiExpQMCLVSO}
\end{center}
\end{figure}

\begin{figure}[t]
 \begin{center}
  \leavevmode
  \epsfxsize=1.0\columnwidth \epsfbox{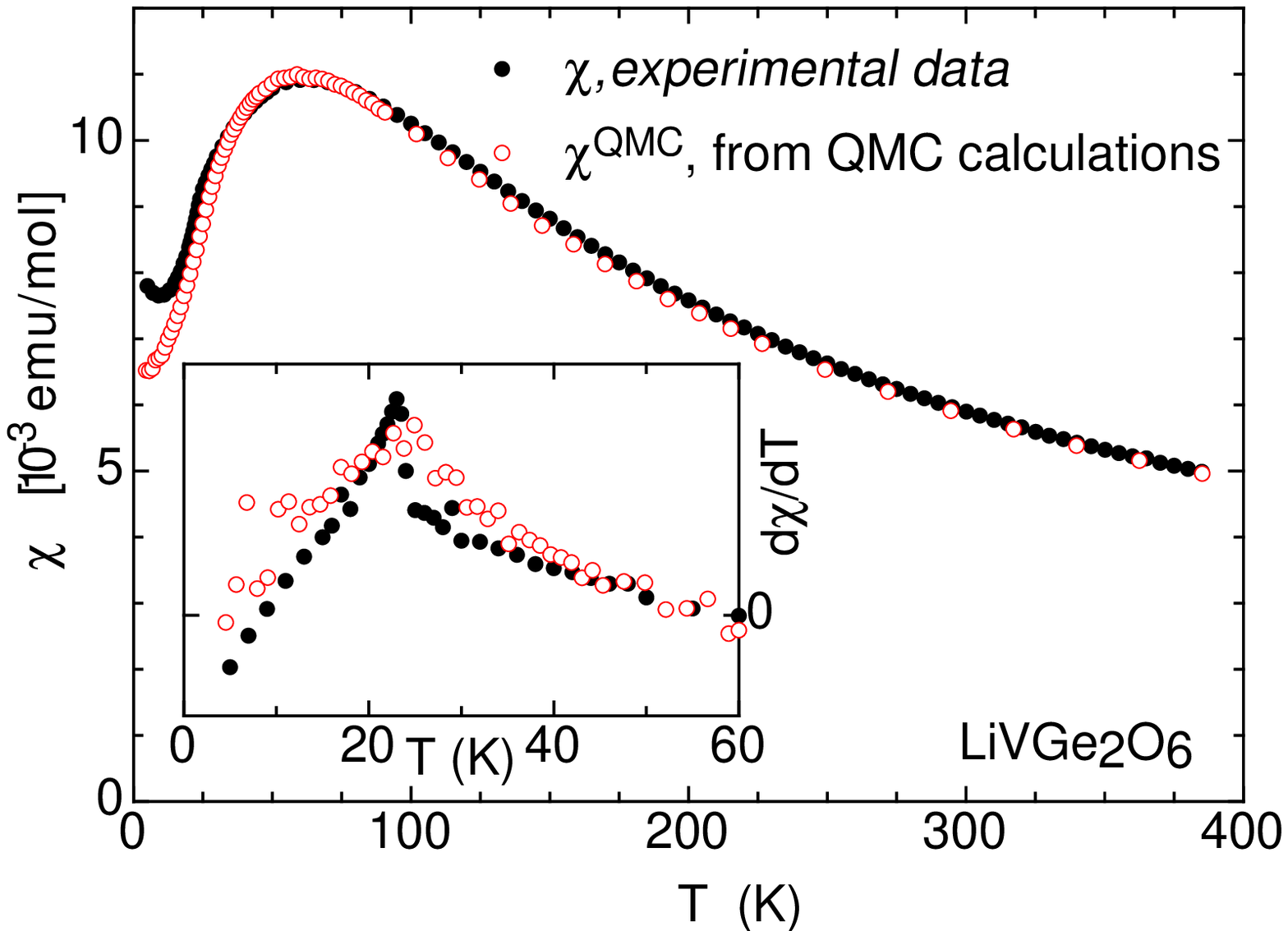}
\caption{
Magnetic susceptibility $\chi$ as a function of temperature $T$ of LiVGe$_2$O$_6$, compared with the $\chi(T)$ resulting from quantum Monte Carlo calculations.
Inset: $d\chi/dT(T)$.
}
\label{FigChiExpQMCLVGO}
\end{center}
\end{figure}

\begin{figure}[t]
 \begin{center}
  \leavevmode
  \epsfxsize=1.0\columnwidth \epsfbox{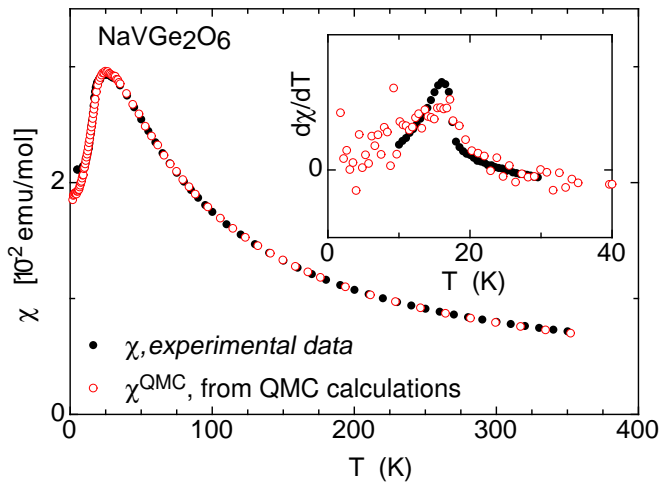}
\caption{
Magnetic susceptibility $\chi$ as a function of temperature $T$ of NaVGe$_2$O$_6$, compared with the $\chi(T)$ resulting from quantum Monte Carlo calculations.
Inset: $d\chi/dT(T)$.
}
\label{FigChiExpQMCNVGO}
 \end{center}
\end{figure}

Next, we compare the experimental $\chi(T)$ data of LiVGe$_2$O$_6$ \cite{Gavilano_2000_PRL} and NaVGe$_2$O$_6$ \cite{Pedrini_2004_PRB} 
with those calculated with QMC methods, employing the values of $J_{\mathrm{p}}$ and $J$ from Table \ref{TabJpJ}.
This is done in Figs.~\ref{FigChiExpQMCLVSO}, \ref{FigChiExpQMCLVGO}, and \ref{FigChiExpQMCNVGO}, where $\chi$ is plotted as a function of temperature $T$.
Bullets show the experimental data, and
open circles the calculated $\chi^{\mathrm{QMC}}(T)$, adapted to the experimental data by varying the parameters $\chi_0$ and $p_{\mathrm{eff}}$ in
\begin{equation}
  \label{EqChiRescaling}
  \chi^{\mathrm{QMC}}(T)=\chi_0 + (N_{\mathrm{AV}}\mu_{\mathrm{B}}^2)p_{\mathrm{eff}}^2\;J\bar{\chi}(t=k_{\mathrm{B}}T/J)
  \quad.
\end{equation}
In Eq.~(\ref{EqChiRescaling}) $\bar{\chi}$ and $t$ are the dimensionless susceptibility and temperature, respectively, discussed in Appendix \ref{AppSusc}.
Their relations to physical quantities is exemplified in  Table~\ref{TabRescaling}.
The factor $(N_{\mathrm{AV}}\mu_{\mathrm{B}}^2)$ accounts for the correct units of a magnetic susceptibility.
The additive constant $\chi_0$  represents the offset in the experimental data due to diamagnetic contributions and $p_{\mathrm{eff}}^2$ is a multiplicative factor, causing the high-temperature magnetic susceptibility  
$(N_{\mathrm{AV}}\mu_{\mathrm{B}}^2)p_{\mathrm{eff}}^2S(S+1)/3k_{\mathrm{B}}T$ to be of Curie-Weiss type.
We note that the value of $p_{\mathrm{eff}}$ of NaVGe$_2$O$_6$ established in this way is somewhat larger than the one which resulted from the reported fitting of a Curie-Weiss law to the experimental $\chi(T)$ data at high temperatures  \cite{Pedrini_2004_PRB}.
The insets of Figs.~\ref{FigChiExpQMCLVSO}, \ref{FigChiExpQMCLVGO}, and \ref{FigChiExpQMCNVGO} represent the derivatives with respect to temperature, 
$(d\chi/dT)(T)$, and $(d\chi^{\mathrm{QMC}}/dT)(T)$.
The discrepancy between the two can only partially be attributed to the presence of paramagnetic impurities, 
which might reduce $d\chi/dT$  at low temperatures.
The prominent peak in $d\chi/dT$ for LiVGe$_2$O$_6$ and NaVGe$_2$O$_6$ 
(less so for LiVSi$_2$O$_6$) is not accounted for by the simulations, 
indicating that the Hamiltonian in Eq.~(\ref{EqHamilton}) is not able to fully account for the magnetic properties of the AVX$_2$O$_6$-series.

A more realistic model would include a term
\begin{equation}
  \label{EqHamiltonD}
   H_D = -D\sum_{a;i}S^z_{a,i} S^z_{a,i}
   \quad,
\end{equation}
in the Hamiltonian of Eq. (\ref{EqHamilton}), which describes a single ion magnetic anisotropy. 
$D>0$ favorizes the alignement of the moments along the $z$-axis, while $D<0$ favorizes configurations 
with magnetic moments in the plane perpendicular to the $z$-axis.
As discussed in the Appendix, a term with $D>0$ is appropriate for the present case.
Indeed, it is the natural choice in order to account for the formation of the gap $\Delta$ in the magnetic excitation spectrum, and to reproduce the sharp peak in $d\chi/dT$.
This possibility was already considered in Ref. \cite{Vonlanthen_2002_PRB}.
Further speculations would require a more detailed analysis of weakly coupled anisotropic spin $S=1$ chains, 
which we leave for future investigations.
Here, we only remark that Eqs. (\ref{EqJpJTNJ}) and (\ref{EqJpJTmax}) have to be adapted if $D\neq0$, which would yield values of $J_{\mathrm{p}}$ and $J$ different from those given in Table \ref{TabJpJ}.

\section{\label{SecSummary}Summary, Conclusions and Outlook}

Our experimental investigations confirm that in the AVX$_2$O$_6$ compounds the interchain coupling $J_{\mathrm{p}}$ is always large enough to quench the Haldane gap and provoke the onset of a three-dimensional antiferromagnetic order at $T_{\mathrm{N}}>0$.
While the calculated $J_{\mathrm{p}}$ is almost constant across the AVX$_2$O$_6$ series, the intrachain coupling $J$ varies by a factor of 5.
This variation does not seem to be correlated to geometrical parameters of the V chains.
We therefore suspect that the electronic orbitals that are involved in the mediation of the 
coupling between the V magnetic moments along the chains, are affected by the positions of the other atoms in the structure.
In particular, the growing distortion of the VO$_6$ octahedra seems to reduce the value of $J$.
Band structure calculations and the subsequent determination of the Wannier functions of the V and O atoms of the four different compounds have to be made in order to understand the details of the intrachain V-V coupling.

The comparison of the experimental and theoretical temperature dependences of the magnetic susceptibility indicates that the magnetic moments at the V atoms are affected by magnetic easy axis single-ion anisotropy, 
consistent with the observation of a gap in the magnetic excitation spectra below $T_{\mathrm{N}}$.
The origin and the magnitude of this effect could also be clarified starting from the results of band structure calculations.

\begin{acknowledgements}
We are grateful to M. Sigrist and R. Monnier for useful discussions.
We acknowledge the help of M. Weller in the preparation of the manuscript.
The numerical simulations were performed using the Asgard and Hreidar cluster at ETH Z\"urich.
\end{acknowledgements}

\begin{appendix}

\section{\label{AppSusc}QMC calculations, Magnetic Susceptibility and Critical Temperature}

In this Appendix we focus on the QMC calculations, aiming at clarifying some points of the discussion of 
the magnetic susceptibility in the AVX$_2$O$_6$ series in Sec.~\ref{SecJpJ}.
For the QMC computations, the intrachain coupling is set to unity, i.e., $J=1$, and all quantities are dimensionless.
We employ this convention throughout this Appendix.
Physical quantities with correct dimensions can be gained by using the corresponding energy values of $J$, according to Table \ref{TabRescaling}.

\begin{table}[t]
 \begin{center}
 \begin{tabular}{|c|c|}
   \hline
    &  \\
    Dimensionless & Physical\\
    quantity      & quantity \\
    &  \\
    \hline
    &  \\
    1 & $J$ \\
    $t$ & $T=J/k_B\cdot t$ \\
    $j_{\mathrm{p}}$ & $J_{\mathrm{p}}=J\cdot j_{\mathrm{p}}$ \\
    $\bar{\chi}$ & $\chi=J\cdot(N_{\mathrm{AV}}\mu_{\mathrm{B}}^2)\bar{\chi}$ \\
    $\delta$ & $\Delta=J\cdot\delta$ \\
    $d$ & $D=J\cdot d$ \\
    $h$ & $H=J\cdot h$ \\
    $c$ & $C=c\cdot N_{\mathrm{AV}}k_B$ \\
    &  \\
    \hline
 \end{tabular}
\caption{
Conversion from dimensionless to physical quantities.}
\label{TabRescaling}
\end{center}
\end{table}

The QMC calculations are based on the Hamiltonian
\begin{equation}
  \label{Eqhamilton}
  h = h_j+j_{j_{\mathrm{p}}}+h_d
  \quad,
\end{equation}
where the three terms are the intrachain Hamiltonian (see also Eq. (\ref{EqHamiltonJ}))
\begin{equation}
  \label{Eqhamiltonj}
   h_j = j\sum_{r;i}\vec{S}_{r,i}\cdot\vec{S}_{r,i+1}
   \quad (j>0) \quad,
\end{equation}
the interchain Hamiltonian (see also Eq. (\ref{EqHamiltonJp}))
\begin{equation}
  \label{Eqhamiltonjp}
   h_{j_{\mathrm{p}}} = j_{\mathrm{p}}\sum_{(r,s);i}\vec{S}_{r,i}\cdot\vec{S}_{s,i}
   \quad,
\end{equation}
and the magnetic anisotropy hamiltonian (see also (\ref{EqHamiltonD}))
\begin{equation}
  \label{Eqhamiltonjd}
   h_d = -d\sum_{r;i}S^z_{r,i} S^z_{r,i}
   \quad.
\end{equation}
For noninteracting chains ($j_{\mathrm{p}}=0$), 
we considered a one-dimensional array of up to 100 $S=1$ spins.
For weakly interacting chains we used a cubic $8\times8\times20$-lattice of $S=1$ spins, with the largest extension corresponding to the chain direction.
The spin  $S=1$ operators of the $i$-th spin on chain $r$ are denoted by $\vec{S}_{r,i}$.
With $(r,s)$ we denote a pair of neighbouring chains.
We employed a stochastic series expansion~\cite{Sandvik_1999_PRB,Alet_2005_PRE} 
QMC code based on the ALPS library~\cite{ALPS,Troyer_1998_LNCS}.

\begin{figure}[t]
 \begin{center}
  \leavevmode
  \epsfxsize=1.0\columnwidth \epsfbox{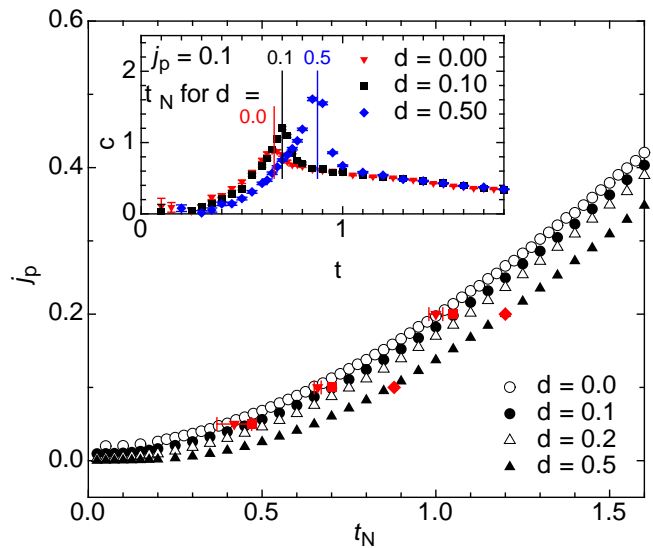}
\caption{
Interchain coupling $j_{\mathrm{p}}$ as a function of $t_{\mathrm{N}}$, determined using the modified RPA according to Eq. (\ref{EqjptN})  (see also 
\cite{Yasuda_2005_PRL,Pedrini_2004_PRB}), for different values of $d>0$.
The red symbols represent points $(t_{\mathrm{N}},j_{\mathrm{p}})$, 
where $t_{\mathrm{N}}$ is the temperature of the maximum specific heat $c$ for a system with interchain coupling $j_{\mathrm{p}}$.
Inset: $c_p(t)$ for $j_{\mathrm{p}}=0.1$ and different values of $d\geq0$, calculated with QMC methods; the critical temperature $t_{\mathrm{N}}$ is indicated.
}
\label{FigjptN}
\end{center}
\end{figure}

In the RPA approximation the effect of the interchain interaction $h_{j_{\mathrm{p}}}$ is treated as an external perturbation of a single chain by the $n$ 
neighbouring chains.
Standard mean field arguments show that the temperature $t_{\mathrm{N}}$ 
of the transition from the high-temperature paramagnetic to the low-temperature antiferromagnetically ordered state is related to the 
interchain coupling according to 
\begin{equation}
  \label{EqjptN}
   j_{\mathrm{p}}=\frac{1}{\xi N \bar{\chi}^s_{zz}(t_{\mathrm{N}})}
   \quad.
\end{equation}
The staggered susceptibility $\bar{\chi}^s_{zz}$ per spin of an isolated $S=1$ chain, described by the hamiltonian $h_j+h_d$, was calculated using the 
QMC technique for various values $0\leq d\leq0.5$.
By analyzing the results of QMC calculations for the 
Hamiltonian (\ref{Eqhamilton}) on a cubic lattice ($n=4$) with $d=0$, it was shown in Ref.~\cite{Yasuda_2005_PRL} that, for $j_{\mathrm{p}}\leq0.2$, the 
interchain coupling renormalizes the value of $\xi$ in Eq. (\ref{EqjptN}) from unity down to $\xi=0.695$.
In Fig.~\ref{FigjptN} we display the result for $j_{\mathrm{p}}(t_{\mathrm{N}})$ also for the case $d\geq0$
\footnote{
From our analysis it follows that in a single chain, the Haldane gap 
is quenched if $d$ exceeds a critical value of $d_{\mathrm{crit}}\approx0.5$.
This follows from $j_{\mathrm{p}}\rightarrow0$ for $T\rightarrow0$.
This observation is in good agreement with the results reported in Ref.~\cite{Chen_2003_PRB}.
}.
We find good agreement to discrete pairs of values $(j_{\mathrm{p}},t_{\mathrm{N}})$ 
resulting of QMC computations using the full Hamiltonian (\ref{Eqhamilton}).
Such calculations yield, among other quantities, the temperature dependence of the 
specific heat $c$. For a given value of $j_{\mathrm{p}}$, the temperature of the peak in $c$ 
is identified as $t_{\mathrm{N}}$ (see inset of Fig.~\ref{FigjptN}).

\begin{figure}[t]
 \begin{center}
  \leavevmode
  \epsfxsize=1.0\columnwidth \epsfbox{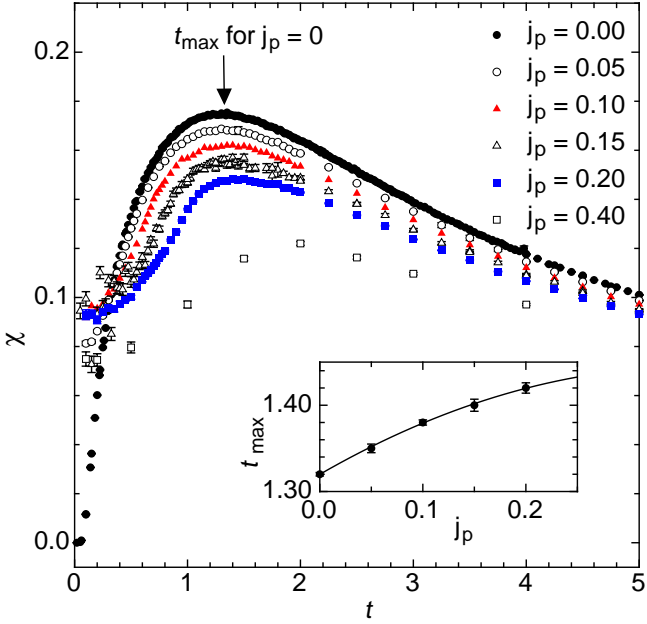}
\caption{
Magnetic susceptibility $\bar{\chi}$ calculated using QMC, as a function of the temperature $t$, for different values of $j_{\mathrm{p}}$. 
For $j_{\mathrm{p}}=0$ the location of $t_{\mathrm{max}}$ is indicated.
Inset: $t_{\mathrm{max}}$ as a function of $j_{\mathrm{p}}$. The solid line represents the best fitting second order polynomial.
}
\label{Figjptmax}
\end{center}
\end{figure}

We first focus on the case $d=0$.
In Fig. \ref{Figjptmax} we show examples of the temperature dependence of the magnetic susceptibility $\bar{\chi}(T)$, calculated using QMC, 
for selected values of $j_{\mathrm{p}}$.
The temperature $t_{\mathrm{max}}$, corresponding to the maximum of $\bar{\chi}$, is indicated.
The inset of Fig. \ref{Figjptmax} displays $t_{\mathrm{max}}(j_{\mathrm{p}})$ which we find to be well approximated by 
\begin{equation}
  \label{Eqjptmax}
  t_{\mathrm{max}}=a_2 j_{\mathrm{p}}^2+a_1 j_{\mathrm{p}} +a_0
  \quad,
\end{equation}
with $a_2=-0.95$, $a_1=0.69$ and $a_0=1.32$.

In this way, one obtains
\begin{equation}
  \label{EqtNtmaxjp}
  \frac{t_{\mathrm{N}}}{t_{\mathrm{max}}}(j_{\mathrm{p}})=
  \frac{t_{\mathrm{N}}(j_{\mathrm{p}})}{t_{\mathrm{max}}(j_{\mathrm{p}})}
  \quad,
\end{equation}
where the numerator of the right hand side is the inverse function of $j_{\mathrm{p}}(t_{\mathrm{N}})$ given in Eq.~(\ref{EqjptN}), and the denominator is given 
in Eq.~(\ref{Eqjptmax}).
The inverse of the function in Eq.~(\ref{EqtNtmaxjp}), 
\begin{equation}
  \label{EqjptNtmax}
  j_{\mathrm{p}}(t_{\mathrm{N}}/t_{\mathrm{max}})
  \quad,
\end{equation}
is plotted in Figure \ref{FigJpJTNTmax} in the form $(J_{\mathrm{p}}/J)(T_{\mathrm{N}}/T_{\mathrm{max}})$.

\begin{figure}[t]
 \begin{center}
  \leavevmode
  \epsfxsize=1.0\columnwidth \epsfbox{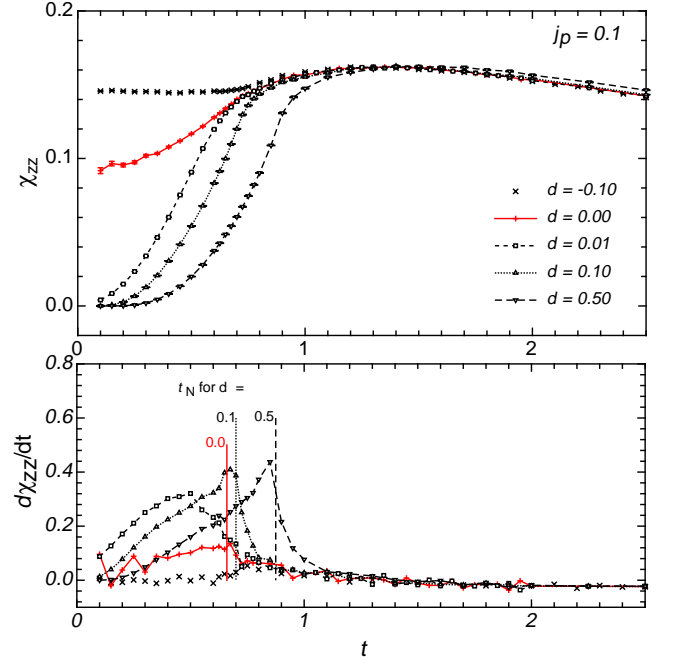}
\caption{
$\bar{\chi}_{zz}(t)$ (upper plot) and $d\bar{\chi}_{zz}/dt(t)$ (lower plot), determined with QMC calculations, for fixed $j_{\mathrm{p}}=0.1$ and different values of $d$.
The transition temperature $t_{\mathrm{N}}$, corresponding to the peak in $c(t)$, is indicated.
}
\label{Figchitjp01}
\end{center}
\end{figure}

Finally we consider the case of $d\neq0$ which introduces an anisotropy of the magnetic susceptibility 
such that $\bar{\chi}_{zz}\neq\bar{\chi}_{xx}=\bar{\chi}_{yy}$.
Averaging over all directions, the effective value for a powder sample is
\begin{equation}
  \label{Eqchipowder}
  \bar{\chi}_{\mathrm{powder}}=\frac{1}{3}\bar{\chi}_{zz}+\frac{2}{3}\bar{\chi}_{xx}
  \quad.
\end{equation}
The temperature dependence of $\bar{\chi}_{zz}(t)$ has been computed by QMC 
and is represented in the upper panel of Fig. \ref{Figchitjp01} for $j_{\mathrm{p}}=0.1$ and selected values of $d$.
Note that Eq. (\ref{Eqjptmax}), and thus also (\ref{EqjptNtmax}), do not hold if $d>0$, 
since for fixed $j_{\mathrm{p}}$, the 
maximum of $\bar{\chi}_{zz}(t)$ is shifted towards higher temperatures.
Nevertheless, 
two important conclusions can be drawn from the behaviour of $\bar{\chi}_{zz}(t)$:
First, $d>0$ induces an anomaly in $\bar{\chi}_{zz}(t)$ at $t_{\mathrm{N}}$, as is shown in the lower panel of Fig. \ref{Figchitjp01}.
This kink will be reflected in $\bar{\chi}_{\mathrm{powder}}$ (see Eq. (\ref{Eqchipowder})).
Secondly, for $d>0$, the rapid decay of $\bar{\chi}_{zz}$ to zero upon cooling below $t_{\mathrm{N}}$ indicates the formation of an energy gap $\delta$ above the three-dimensional antiferromagnetically ordered ground state (in agreement with Ref.~\cite{Hone_1969_PR}).
The low-temperature behaviour is consistent with an exponential suppression,
\begin{equation}
  \label{EqChiLowT}
  \bar{\chi}_{zz}(t)\sim\exp(-\delta/t)
  \quad,
\end{equation}
indicative of a gap $\delta$ in the system.
We leave a detailed analyis of the dependence of $\delta$ 
on $d$ and $j_{\mathrm{p}}$ to future investigations.
Finally, we note that
for $d<0$, the presence of a gap $\delta$ is excluded and
in this case, $\bar{\chi}_{zz}$ tends to a non-zero value for $t\rightarrow0$.

\end{appendix}

\bibliography{LVSO}

\begin{thebibliography}{19}
\expandafter\ifx\csname natexlab\endcsname\relax\def\natexlab#1{#1}\fi
\expandafter\ifx\csname bibnamefont\endcsname\relax
  \def\bibnamefont#1{#1}\fi
\expandafter\ifx\csname bibfnamefont\endcsname\relax
  \def\bibfnamefont#1{#1}\fi
\expandafter\ifx\csname citenamefont\endcsname\relax
  \def\citenamefont#1{#1}\fi
\expandafter\ifx\csname url\endcsname\relax
  \def\url#1{\texttt{#1}}\fi
\expandafter\ifx\csname urlprefix\endcsname\relax\def\urlprefix{URL }\fi
\providecommand{\bibinfo}[2]{#2}
\providecommand{\eprint}[2][]{\url{#2}}

\bibitem[{\citenamefont{Haldane}(1983)}]{Haldane_1983_PhysLett}
\bibinfo{author}{\bibfnamefont{F.~D.~M.} \bibnamefont{Haldane}},
  \bibinfo{journal}{Phys.\ Lett.} \textbf{\bibinfo{volume}{93A}},
  \bibinfo{pages}{464} (\bibinfo{year}{1983}).

\bibitem[{\citenamefont{Millet et~al.}(1999)\citenamefont{Millet, Mila, Zhang,
  Mambrini, Van-Osten, Paschenko, Sulpice, and Stepanov}}]{Millet_1999_PRL}
\bibinfo{author}{\bibfnamefont{P.}~\bibnamefont{Millet}},
  \bibinfo{author}{\bibfnamefont{F.}~\bibnamefont{Mila}},
  \bibinfo{author}{\bibfnamefont{F.~C.} \bibnamefont{Zhang}},
  \bibinfo{author}{\bibfnamefont{M.}~\bibnamefont{Mambrini}},
  \bibinfo{author}{\bibfnamefont{A.~B.} \bibnamefont{Van-Osten}},
  \bibinfo{author}{\bibfnamefont{A.}~\bibnamefont{Paschenko}},
  \bibinfo{author}{\bibfnamefont{A.}~\bibnamefont{Sulpice}}, \bibnamefont{and}
  \bibinfo{author}{\bibfnamefont{A.}~\bibnamefont{Stepanov}},
  \bibinfo{journal}{Phys.\ Rev.\ Lett.} \textbf{\bibinfo{volume}{83}},
  \bibinfo{pages}{4176} (\bibinfo{year}{1999}).

\bibitem[{\citenamefont{Gavilano et~al.}(2000)\citenamefont{Gavilano,
  Mushkolaj, Ott, Millet, and Mila}}]{Gavilano_2000_PRL}
\bibinfo{author}{\bibfnamefont{J.~L.} \bibnamefont{Gavilano}},
  \bibinfo{author}{\bibfnamefont{S.}~\bibnamefont{Mushkolaj}},
  \bibinfo{author}{\bibfnamefont{H.~R.} \bibnamefont{Ott}},
  \bibinfo{author}{\bibfnamefont{P.}~\bibnamefont{Millet}}, \bibnamefont{and}
  \bibinfo{author}{\bibfnamefont{F.}~\bibnamefont{Mila}},
  \bibinfo{journal}{Phys.\ Rev.\ Lett.} \textbf{\bibinfo{volume}{85}},
  \bibinfo{pages}{409} (\bibinfo{year}{2000}).

\bibitem[{\citenamefont{Vonlanthen et~al.}(2002)\citenamefont{Vonlanthen,
  Tanaka, Goto, Clark, Millet, Henry, Gavilano, Ott, Mila, Berthier
  et~al.}}]{Vonlanthen_2002_PRB}
\bibinfo{author}{\bibfnamefont{P.}~\bibnamefont{Vonlanthen}},
  \bibinfo{author}{\bibfnamefont{K.~B.} \bibnamefont{Tanaka}},
  \bibinfo{author}{\bibfnamefont{A.}~\bibnamefont{Goto}},
  \bibinfo{author}{\bibfnamefont{W.~G.} \bibnamefont{Clark}},
  \bibinfo{author}{\bibfnamefont{P.}~\bibnamefont{Millet}},
  \bibinfo{author}{\bibfnamefont{J.~Y.} \bibnamefont{Henry}},
  \bibinfo{author}{\bibfnamefont{J.~L.} \bibnamefont{Gavilano}},
  \bibinfo{author}{\bibfnamefont{H.~R.} \bibnamefont{Ott}},
  \bibinfo{author}{\bibfnamefont{F.}~\bibnamefont{Mila}},
  \bibinfo{author}{\bibfnamefont{C.}~\bibnamefont{Berthier}},
  \bibnamefont{et~al.}, \bibinfo{journal}{Phys.\ Rev.\ B}
  \textbf{\bibinfo{volume}{65}}, \bibinfo{pages}{214413}
  (\bibinfo{year}{2002}).

\bibitem[{\citenamefont{Pedrini et~al.}(2004)\citenamefont{Pedrini, Gavilano,
  Rau, Ott, Karpinski, and Wessel}}]{Pedrini_2004_PRB}
\bibinfo{author}{\bibfnamefont{B.}~\bibnamefont{Pedrini}},
  \bibinfo{author}{\bibfnamefont{J.~L.} \bibnamefont{Gavilano}},
  \bibinfo{author}{\bibfnamefont{D.}~\bibnamefont{Rau}},
  \bibinfo{author}{\bibfnamefont{H.~R.} \bibnamefont{Ott}},
  \bibinfo{author}{\bibfnamefont{J.}~\bibnamefont{Karpinski}},
  \bibnamefont{and} \bibinfo{author}{\bibfnamefont{S.}~\bibnamefont{Wessel}},
  \bibinfo{journal}{Phys.\ Rev.\ B} \textbf{\bibinfo{volume}{70}},
  \bibinfo{pages}{024421} (\bibinfo{year}{2004}).

\bibitem[{\citenamefont{Vasiliev et~al.}(2004)\citenamefont{Vasiliev,
  Ignatchik, Isobe, and Ueda}}]{Vasiliev_2004_PRB}
\bibinfo{author}{\bibfnamefont{A.~N.} \bibnamefont{Vasiliev}},
  \bibinfo{author}{\bibfnamefont{O.~L.} \bibnamefont{Ignatchik}},
  \bibinfo{author}{\bibfnamefont{M.}~\bibnamefont{Isobe}}, \bibnamefont{and}
  \bibinfo{author}{\bibfnamefont{Y.}~\bibnamefont{Ueda}},
  \bibinfo{journal}{Phys.\ Rev.\ B} \textbf{\bibinfo{volume}{70}},
  \bibinfo{pages}{132415} (\bibinfo{year}{2004}).

\bibitem[{\citenamefont{Satto et~al.}(1997)\citenamefont{Satto, Millet, and
  Galy}}]{Satto_1997_ActaCrystC}
\bibinfo{author}{\bibfnamefont{C.}~\bibnamefont{Satto}},
  \bibinfo{author}{\bibfnamefont{P.}~\bibnamefont{Millet}}, \bibnamefont{and}
  \bibinfo{author}{\bibfnamefont{J.}~\bibnamefont{Galy}},
  \bibinfo{journal}{Acta Cryst. C} \textbf{\bibinfo{volume}{53}},
  \bibinfo{pages}{1727} (\bibinfo{year}{1997}).

\bibitem[{\citenamefont{Carter et~al.}(1977)\citenamefont{Carter, Bennett, and
  Kahan}}]{Carter}
\bibinfo{author}{\bibfnamefont{G.~C.} \bibnamefont{Carter}},
  \bibinfo{author}{\bibfnamefont{L.~H.} \bibnamefont{Bennett}},
  \bibnamefont{and} \bibinfo{author}{\bibfnamefont{D.~J.} \bibnamefont{Kahan}},
  \emph{\bibinfo{title}{Metallic Shifts in NMR}} (\bibinfo{publisher}{Pergamon
  Press}, \bibinfo{address}{Oxford}, \bibinfo{year}{1977}).

\bibitem[{\citenamefont{Barak et~al.}(1974)\citenamefont{Barak, Gabai, and
  Kaplan}}]{Barak_1974_PRB}
\bibinfo{author}{\bibfnamefont{J.}~\bibnamefont{Barak}},
  \bibinfo{author}{\bibfnamefont{A.}~\bibnamefont{Gabai}}, \bibnamefont{and}
  \bibinfo{author}{\bibfnamefont{N.}~\bibnamefont{Kaplan}},
  \bibinfo{journal}{Phys.\ Rev.\ B} \textbf{\bibinfo{volume}{9}},
  \bibinfo{pages}{4914} (\bibinfo{year}{1974}).

\bibitem[{\citenamefont{Chen et~al.}(1969)\citenamefont{Chen, Yu, and
  Su}}]{Hone_1969_PR}
\bibinfo{author}{\bibfnamefont{H.}~\bibnamefont{Chen}},
  \bibinfo{author}{\bibfnamefont{L.}~\bibnamefont{Yu}}, \bibnamefont{and}
  \bibinfo{author}{\bibfnamefont{Z.}~\bibnamefont{Su}},
  \bibinfo{journal}{Phys.\ Rev.} \textbf{\bibinfo{volume}{186}},
  \bibinfo{pages}{291} (\bibinfo{year}{1969}).

\bibitem[{\citenamefont{Koga and Kawakami}(2000)}]{Koga_2000_PRB}
\bibinfo{author}{\bibfnamefont{A.}~\bibnamefont{Koga}} \bibnamefont{and}
  \bibinfo{author}{\bibfnamefont{N.}~\bibnamefont{Kawakami}},
  \bibinfo{journal}{Phys.\ Rev.\ B} \textbf{\bibinfo{volume}{61}},
  \bibinfo{pages}{6133} (\bibinfo{year}{2000}).

\bibitem[{\citenamefont{Kim and Birgenau}(2000)}]{Kim_2000_PRB}
\bibinfo{author}{\bibfnamefont{Y.~J.} \bibnamefont{Kim}} \bibnamefont{and}
  \bibinfo{author}{\bibfnamefont{R.~J.} \bibnamefont{Birgenau}},
  \bibinfo{journal}{Phys.\ Rev.\ B} \textbf{\bibinfo{volume}{62}},
  \bibinfo{pages}{6378} (\bibinfo{year}{2000}).

\bibitem[{\citenamefont{Yasuda et~al.}(2005)\citenamefont{Yasuda, Todo,
  Hukushima, Alet, Keller, Troyer, and Takayama}}]{Yasuda_2005_PRL}
\bibinfo{author}{\bibfnamefont{C.}~\bibnamefont{Yasuda}},
  \bibinfo{author}{\bibfnamefont{S.}~\bibnamefont{Todo}},
  \bibinfo{author}{\bibfnamefont{K.}~\bibnamefont{Hukushima}},
  \bibinfo{author}{\bibfnamefont{F.}~\bibnamefont{Alet}},
  \bibinfo{author}{\bibfnamefont{M.}~\bibnamefont{Keller}},
  \bibinfo{author}{\bibfnamefont{M.}~\bibnamefont{Troyer}}, \bibnamefont{and}
  \bibinfo{author}{\bibfnamefont{H.}~\bibnamefont{Takayama}},
  \bibinfo{journal}{Phys.\ Rev.\ Lett.} \textbf{\bibinfo{volume}{94}},
  \bibinfo{pages}{217201} (\bibinfo{year}{2005}).

\bibitem[{\citenamefont{Schulz}(1996)}]{Schulz_1996_PRL}
\bibinfo{author}{\bibfnamefont{H.~J.} \bibnamefont{Schulz}},
  \bibinfo{journal}{Phys.\ Rev.\ Lett.} \textbf{\bibinfo{volume}{77}},
  \bibinfo{pages}{2790} (\bibinfo{year}{1996}).

\bibitem[{\citenamefont{Sandvik}(1999)}]{Sandvik_1999_PRB}
\bibinfo{author}{\bibfnamefont{A.~W.} \bibnamefont{Sandvik}},
  \bibinfo{journal}{Phys.\ Rev.\ B} \textbf{\bibinfo{volume}{59}},
  \bibinfo{pages}{R14157} (\bibinfo{year}{1999}).

\bibitem[{\citenamefont{Alet et~al.}(1999)\citenamefont{Alet, Wessel, and
  Troyer}}]{Alet_2005_PRE}
\bibinfo{author}{\bibfnamefont{F.}~\bibnamefont{Alet}},
  \bibinfo{author}{\bibfnamefont{S.}~\bibnamefont{Wessel}}, \bibnamefont{and}
  \bibinfo{author}{\bibfnamefont{M.}~\bibnamefont{Troyer}},
  \bibinfo{journal}{Phys.\ Rev.\ E} \textbf{\bibinfo{volume}{71}},
  \bibinfo{pages}{036706} (\bibinfo{year}{1999}).

\bibitem[{\citenamefont{Alet}(12005)}]{ALPS}
\bibinfo{author}{\bibfnamefont{F.}~\bibnamefont{Alet}},
  \bibinfo{journal}{Phys.\ Soc.\ Jpn.\ Suppl.} \textbf{\bibinfo{volume}{74}},
  \bibinfo{pages}{30} (\bibinfo{year}{12005}).

\bibitem[{\citenamefont{Troyer$\;$\textit{et al.}}(1998)}]{Troyer_1998_LNCS}
\bibinfo{author}{\bibfnamefont{M.}~\bibnamefont{Troyer$\;$\textit{et al.}}},
  \bibinfo{journal}{Lect.\ Notes Comput.\ Sci.}
  \textbf{\bibinfo{volume}{1505}}, \bibinfo{pages}{191} (\bibinfo{year}{1998}).

\bibitem[{\citenamefont{Chen et~al.}(2003)\citenamefont{Chen, Hida, and
  Sanctuary}}]{Chen_2003_PRB}
\bibinfo{author}{\bibfnamefont{W.}~\bibnamefont{Chen}},
  \bibinfo{author}{\bibfnamefont{K.}~\bibnamefont{Hida}}, \bibnamefont{and}
  \bibinfo{author}{\bibfnamefont{B.~C.} \bibnamefont{Sanctuary}},
  \bibinfo{journal}{Phys.\ Rev.\ B} \textbf{\bibinfo{volume}{67}},
  \bibinfo{pages}{104401} (\bibinfo{year}{2003}).

\end{thebibliography}

\end{document}